\documentclass[
showpacs,
twocolumn
,prl,aps,floatfix]{revtex4-1}
\pdfoutput=1
\usepackage{graphicx}
\usepackage[centertags]{amsmath}
\usepackage{amsfonts} 
\usepackage{amssymb}

\begin{document}
\title{More constraining conformal bootstrap}
\author{Ferdinando Gliozzi$~^{a,b}$ }
\affiliation{$~^a$ School of Computing and Mathematics \& Centre for Mathematical Science, Plymouth University,
Plymouth PL4 8AA, UK\\
 $~^b$ INFN,
Sezione di Torino, via P. Giuria, 1, I-10125 Torino, Italy}
\newcommand{\eq}{\begin{equation}}
\newcommand{\en}{\end{equation}}
\newcommand{\ear}{\begin{eqnarray}}
\newcommand{\rae}{\end{eqnarray}}
\newcommand{\N}{\mathbb{N}}
\newcommand{\Z}{\mathbb{Z}}
\newcommand{\C}{\mathbb{C}}
\newcommand{\uu}{\mathbb{I}}
\newcommand{\R}{{\cal R}}
\newcommand{\s}{{\cal S}}
\newcommand{\T}{{\cal T}}
\newcommand{\D}{{\cal D}}
\newcommand{\OO}{{\cal O}}
\newcommand{\oto}{\leftrightarrow}
\newcommand{\bra}{\langle}
\newcommand{\ket}{\rangle}
\newcommand{\um}{\frac12}

\begin{abstract}
Recently an efficient numerical method has been developed to implement
the constraints of crossing symmetry and unitarity on the operator dimensions and OPE coefficients of conformal field theories (CFT) in diverse space-time 
dimensions. It appears that the calculations can  be done only for theories lying at the boundary of the allowed parameter space. Here it is pointed out that a similar method can be applied to a larger class of CFT's, whether unitary or not, and no free parameter remains, provided we know the fusion 
algebra of the low lying primary operators.  As an example we calculate 
using  first principles, with no phenomenological input, the lowest scaling dimensions 
of the local operators associated with the Yang-Lee edge singularity in three 
and four space dimensions. The edge exponents compare favorably with the latest 
numerical estimates. A consistency check of this approach on the 3d critical 
Ising model is also made.     

\end{abstract}
\maketitle

One of the manifold expressions of the bootstrap dream is the conformal 
bootstrap, i.e. the idea that the crossing symmetry of the four-point functions of a CFT is so constraining 
\cite{{Ferrara:1973yt},{Polyakov:1974gs},{Polyakov:1984yq}}   
that, in some cases, it could uniquely fix  the spectrum of the allowed 
scaling dimensions of the theory.   
In the case of two-dimensional rational CFT's, i.e. those with a finite number of Virasoro primary fields, there is an almost perfect 
implementation of this idea encoded in the Vafa equations  \cite{Vafa:1988ag}. These are Diophantine equations built by combining the crossing symmetry with the surprising modular properties of the fusion algebra \cite{Verlinde:1988sn}. As a result the spectrum 
of Virasoro primary fields turns out to be discrete and all the scaling 
dimensions are determined modulo an integer. 

The aim of this Letter is to reformulate the numerical  method recently developed to implement the constraints of crossing symmetry in CFT's in diverse space-time dimensions \cite{{Rattazzi:2008pe},{ElShowk:2012ht},{ElShowk:2012hu}} so that it resembles, to a certain degree, Vafa equations, provided rational CFT's in two dimensions are replaced  
with  truncable CFT's in  $D$-dimensions, i.e theories whose four-point functions can be well approximated by a finite sum of conformal blocks (in a way to be specified later); the Diophantine equations are replaced by transcendental ones whose solutions yield  approximate values of the scaling dimensions of the 
primary operators of the theory. The level of the approximation is controlled by the number of conformal blocks considered. 

As in Vafa equations, unitarity plays no 
role in this reformulation. While unitarity should be invoked for a sensible quantum field theory, many interesting critical systems do not correspond to unitary theories, thus it is important to extend the method to these systems.

The starting point of this kind of analysis is a suitable 
parameterization of the  four-point function of a scalar field $\varphi(x)$ in a $D$-dimensional CFT. The $SO(D+1,1)$  conformal invariance makes it possible  
to write 
\eq
\bra\varphi(x_1)\varphi(x_2)\varphi(x_3)\varphi(x_4)\ket=
\frac{g(u,v)}{\vert x_{12}\vert^{2\Delta_\varphi}\vert x_{34}\vert^{2\Delta_\varphi}},
\label{fourpoint}
\en 
where $\Delta_\varphi$ is the scaling dimension of $\varphi$,  $x_{ij}^2$ is the square of the distance between $x_i$ 
and $x_j$, $g(u,v)$ is an arbitrary function of the cross-ratios 
$u=\frac{x_{12}^2x_{34}^2 }{x_{13}^2x_{24}^2 }$ and
 $v=\frac{x_{14}^2x_{23}^2 }{x_{13}^2x_{24}^2 }$. The function $g$ can be expanded in terms of the conformal blocks $G_{\Delta,L}(u,v)$, i.e.  the eigenfunctions of the Casimir operator of $SO(D+1,1)$:
\eq
g(u,v)=1+\sum_{\Delta,L}{\sf p}_{\Delta,L}G_{\Delta,L}(u,v).
\label{bexpansion}
\en
The coefficients ${\sf p}_{\Delta,L}$ determine, up to a sign, the operator 
product expansion (OPE) of $\varphi(x_1)\varphi(x_2)$. Namely,
if ${\sf p}_{\Delta,L} \not=0$, there is in this OPE a primary operator 
$\OO$ of scaling dimension $\Delta$ and spin $L$ with an OPE coefficient 
$\lambda_{\varphi\varphi\OO}$ with 
$\lambda_{\varphi\varphi\OO}^2= {\sf p}_{\Delta,L}$. In the following it is not necessary to know the detailed form of the OPE but simply its fusion rule that, in the case at hand, we write as
\eq
[\Delta_\varphi]\times [\Delta_\varphi]=\sum_i N_i[\Delta_i,L_i], 
\label{fusion}
\en
where the integer $N_i$ denotes the number of different primary operators of quantum numbers $\Delta_i$ and $L_i$ and we set $[\Delta]\equiv[\Delta,0]$.  

The LHS of (\ref{fourpoint}) is  invariant under any permutation of the four coordinates $x_i$, while the RHS is not, unless $g(u,v)$ fulfils two functional equations which express the crossing symmetry constraints. The $x_1\leftrightarrow x_2$ interchange gives $g(u,v)=g(u/v,1/v)$ which is followed only by the conformal  blocks of 
even spin in (\ref{bexpansion}), so only they contribute to 
(\ref{bexpansion}). The  $x_1\leftrightarrow x_3$ interchange 
yields a functional equation that we write in the form used in 
\cite{Rattazzi:2008pe}
\eq
\sum_{\Delta,L}{\sf p}_{\Delta,L}
\frac{v^{\Delta_\varphi}G_{\Delta,L}(u,v)-u^{\Delta_\varphi}G_{\Delta,L}(v,u)}
{u^{\Delta_\varphi}- v^{\Delta_\varphi}}=1\,.
\label{crossing}
\en
Following \cite{Rattazzi:2008pe}   we put $u=z\bar{z}$ and $v=(1-z)(1-\bar{z})$. In Euclidean space $\bar{z}$ is the complex conjugate of $z$ while in Minkowski
space they can be treated as independent  real variables. The conformal blocks are smooth functions in the region $0\leq z,\bar{z}<1$. The central idea of \cite{Rattazzi:2008pe} is  to Taylor expand (\ref{crossing}) about the symmetric point $z=\bar{z}=\um$ and transform  the functional equation into an infinite set of linear equations in the infinite unknowns 
${\sf p}_{\Delta,L}$, made with the derivatives of (\ref{crossing}) of any order. 
To be more specific, 
 following  \cite{ElShowk:2012ht} we make the change of variables 
$z=(a+\sqrt{b})/2$, $\bar{z}=(a-\sqrt{b})/2$
and Taylor expand around $a=1$ and $b=0$. It is easy to see that this expansion will contain only even powers of $(a-1)$ and integer powers of $b$.
The crossing symmetry constraint (\ref{crossing}) can then be rewritten as one inhomogeneous equation
\eq
\sum_{\Delta,L}{\sf p}_{\Delta,L}{\sf f}^{(0,0)}_{\Delta_\varphi,\Delta_L}  =1,
\label{inh}
\en
and an infinite number of homogeneous equations
  \eq
\sum_{\Delta,L}{\sf p}_{\Delta,L}{\sf f}^{(2m,n)}_{\Delta_\varphi,\Delta_L} =0,~
(m,n\in\N,m+n\not=0),
\label{homo}
\en
with
\eq
{\sf f}^{(m,n)}_{\alpha,\beta}=\left(\partial_a^{m}\partial_b^n
\frac{v^{\alpha}G_{\beta}(u,v)-u^{\alpha}G_{\beta}(v,u)}
{u^{\alpha}- v^{\alpha}}\right)_{a,b=1,0}.
\en

At this point our analysis differs from that of \cite{Rattazzi:2008pe}. Instead of looking for a model-independent unitary bound as a consequence of these equations, we  apply the above considerations to some specific CFT, like, 
for instance, a critical $\varphi^4$ theory or a massless free field theory, 
where we assume  the fusion algebra (\ref{fusion}) is known, at least for 
the low-lying primary operators.

We say that this theory is truncable at level $N$ if the partial sum of the first $N$ conformal blocks of the infinite series (\ref{bexpansion}) gives an exact 
solution of a set of $M\ge N$ equations of the homogeneous system (\ref{homo}).
Now a system of $M$ linear homogeneous equations with $N$ unknowns admits a non-identically vanishing solution only if all the minors of order $N$  are 
vanishing \cite{rouche}. This gives rise to $\kappa\leq\left(\begin{matrix}M\cr N\cr\end{matrix}\right)$ independent relations among the scaling dimensions 
$\{\Delta\}_N\equiv[\Delta_\varphi,\Delta,\Delta',\dots ]$ 
of the first $N+1$ primary operators
\eq
{\sf d}_i(\{\Delta\}_N)\equiv
\det[{\sf f}^{(2m_i,n_i)}_{\Delta_\varphi,\Delta\in\{\Delta\}_N}]=0,~(i=1,2\dots\kappa)
\label{minor}
\en
where $m_i,n_i$ indicate the rows belonging to the minor $i$.

At this perturbative order these equations encode the whole amount of information extracted from crossing symmetry, in the sense that the first $M$ homogeneous equations  are exactly solved if and only if all these minors are vanishing
(the inhomogenus equation (\ref{inh}) is simply a normalization condition).
What can they tell us about the physical properties of this truncable CFT? 
The scaling dimension of the energy-momentum tensor 
is fixed to be $\Delta_T=D$ , while we assume initially that the other 
$N$  $\Delta$'s are free parameters. They are progressively constrained by increasing the number $\kappa$ of  equations. The maximum allowed value 
of $\kappa$ in a generic case is  $\kappa=N$, of course, when the system 
(\ref{minor}) has a discrete number of solutions $\{\Delta_{\sf a}\}_N$ 
$({\sf a}=1,2,\dots)$, or even no solution. The latter possibility 
is expected  when one blindly includes in the fusion rule terms which should not be there. On the contrary,  if the 
CFT we are studying truly exists, and we can reasonably infer its fusion 
algebra, we expect  that ${\sf a}\not=0$ and that 
the exact spectrum of the first $N+1$ primary operators $\{\Delta^\star\}_N$ is 
close to one of those discrete solutions (we shall illustrate it with some examples). 

Choosing a (partially) different set of $M$ homogeneous equations, 
the  discrete solutions   $\{\Delta_{\sf a}\}_N$  slightly shift. The extent 
of this displacement gives a measure of the error made in truncating 
the expansion (\ref{bexpansion}) at $N$ conformal blocks.  
It is also  easy to see  that adding a new term in 
the conformal block expansion does not spoil the discrete  
solution  $\{\Delta_{\sf a}\}_N$, but simply induces 
a small correction on it (we leave  the pleasure
of proving it to the interested reader). Thus, if a CFT is truncable 
at level $N$, it is also 
truncable in general at level $N+1$ and so on. 
This fact leads us to conjecture that the truncable CFT's could 
coincide with those with a finite number of primary operators of conformal dimension $\Delta<K$ for any positive $K$. This subclass of CFT's contains many physically interesting examples.      
 
\begin{figure}[tb]
\begin{center}
\includegraphics[width= 8 cm]{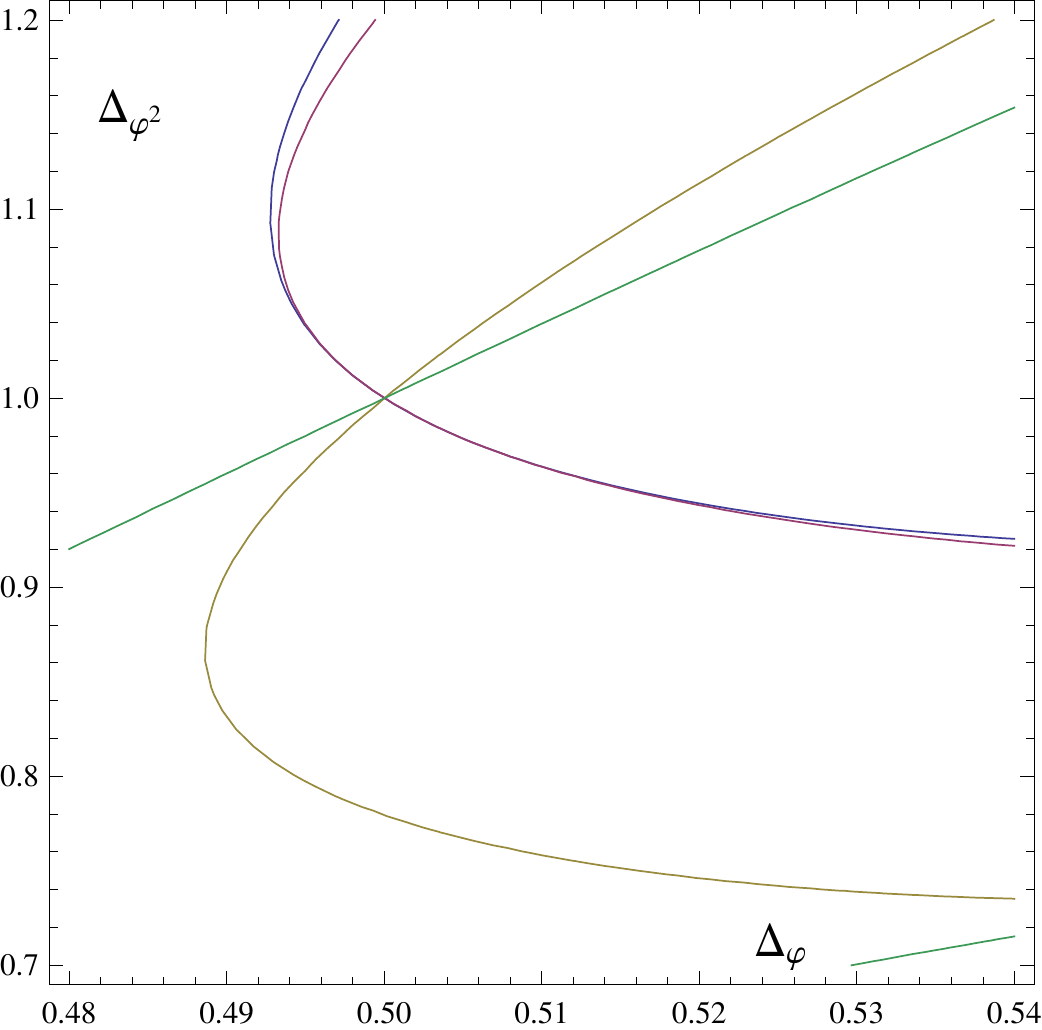}
\end{center}
\caption{Each curve represents the locus of vanishing of a $3\times 3$ minor
of the homogeneous system (\ref{homo}) in the $\Delta_\varphi,\Delta_{\varphi^2}$ plane in the free scalar massless theory in  $D=3$ dimensions.}
\label{Figure:1}  
\end{figure}

As a first application of the present method let us consider  a 
massless free field theory in $D$ space-time dimensions. In this case the fusion rule is 
\eq
[\Delta_\varphi]\times [\Delta_\varphi]=1+[\Delta_{\varphi^2}]+[D,2]+
 [ \Delta_{\varphi^2}+4,4]+\dots,
\label{freefusion}
\en
where  $\Delta_{\varphi^2}=2\Delta_{\varphi}=D-2$,
 but we treat
  $\Delta_{\varphi^2}$ and $\Delta_{\varphi}$ as free parameters and truncate 
(\ref{freefusion}) at $L=4$, resulting in 3 unknowns ${\sf p}_{\Delta,L}$. 
The vanishing of each 
$3\times3$ minor of the homogeneous system gives a relation 
${\sf d}_i(\Delta_\varphi,\Delta_{\varphi^2})=0$ between these two parameters.
Figure \ref{Figure:1} shows four such relations in the $D=3$ case. 
We see that their mutual intersections accumulate around the expected 
exact value. 
Adding now the inhomogeneous equation (\ref{inh}) we obtain the OPE 
coefficients. They accumulate near the exact values given, at $D=3$, by 
the conformal block expansion 
\eq
g(u,v)-1 \equiv \sqrt{u}+\sqrt{\frac uv} =2G_{1,0}+\frac14G_{3,2}+
\frac1{64}G_{5,4}+..
\en
 (We used for the RHS the normalizations of \cite{ElShowk:2012ht}). 

Next, we switch on the interaction  by adding to the action a $\varphi^3$ term with imaginary coupling. Precisely, we put
\eq
S=\int d^Dx\left[\um(\partial\varphi)^2+i(h-h_c)\varphi+ig\varphi^3\right]\,.
\en
This non-unitary theory is known to describe in the infrared  the universality class of the Yang-Lee edge singularity \cite{Fisher:1978pf}. Such a singularity occurs in any 
ferromagnetic $D$-dimensional Ising model above its critical temperature 
$T>T_c$. The zeros of the partition function in the complex plane of the magnetic field $h$ 
are located on the imaginary $ih$ axis  above a critical value $ih_c(T)$. 
In the thermodynamic limit the density of these zeros behaves near $h_c$ like 
$(h-h_c)^\sigma$, where the critical exponent $\sigma$ is related to the 
scaling dimension of the field $\varphi$ by \cite{Fisher:1978pf}
\eq
\sigma=\frac{\Delta_\varphi}{D-\Delta_\varphi}\,.
\label{sigma}
\en
This edge exponent is exactly known in $D=2$ and $D=6$. Our purpose now
is  to evaluate it in $D=3$ and $D=4$ using the present method and to
compare it with 
most recent numerical results. Meanwhile we also check the method in 
$D=2$, where the complete spectrum of primary operators is known as well as  
the OPE coefficients \cite{Cardy:1985yy}.

The $\varphi^3$ interaction tells us that the upper critical dimensionality of this model is $D_u=6$, above which the classical mean-field value $\sigma=\um$ applies. In $6-\epsilon$ there are apparently two relevant operators, $\varphi$
 and  $\varphi^2$; however the latter is, in fact, a redundant operator, as 
at the non-trivial $\varphi^3$ fixed point it is proportional to 
$\partial^2\varphi$ by the equation of motion. Thus  $\varphi^2$ and its derivatives become  descendant operators  of the only relevant primary operator  $\varphi$ of this universality class.  Actually, this is the  only difference between 
the operator content of the Gaussian fixed point of the free-field theory 
and the Yang-Lee edge universality class, as long as the approximate renormalization group analysis applies. As a result, the fusion rule (\ref{freefusion}) becomes even simpler 
\eq
[\Delta_\varphi]\times [\Delta_\varphi]=1+[\Delta_{\varphi}]+[D,2]+
 [ \Delta_4,4]+\dots
\label{ylfusion}
\en
It characterizes the universality class of the Yang-Lee edge 
singularity in any space dimension.

Before inserting such a fusion rule in equations (\ref{minor}), we try to simplify  the notation a bit. Each equation of the homogeneous system (\ref{homo}) is 
labeled by the pair of integers $(m,n)$. We enumerate these equations using the following arbitrary dictionary
\eq
1,2,3,4,5,6,..\oto(1,0)(2,0)(0,1)(0,2)(1,1)(0,3)..
\label{dictionary}
\en  

Let us begin with the $D=2$ case. A $2\times2$ minor of the truncation 
of (\ref{ylfusion}) at $N=2$ 
can be written explicitly in this case as
\eq
{\sf d}_{12}(\Delta_\varphi)=\det\left(\begin{matrix}\partial_a^2G_{\Delta\varphi,0}&
\partial_a^2G_{2,2}\cr\partial_a^4G_{\Delta\varphi,0}&
\partial_a^4G_{2,2}\cr\end{matrix}\right).
\en
It has  a zero at $\Delta_\varphi\simeq-0.422$. Similarly  
${\sf d}_{13}(\Delta_\varphi)$ vanishes at $\Delta_\varphi\simeq-0.362$. The exact value is at $\Delta_\varphi=-\frac25$.  

In order to obtain more accurate results, we have to add the next term of the fusion rule (\ref{ylfusion}), namely the spin 4 operator $[\Delta_4,4]$, 
which depends on the new ``free'' parameter $\Delta_4$. 

In view of the fact that in the fusion rule of any scalar operator in $D=2$ 
the energy momentum tensor $T+\bar{T}$  is always accompanied by the scalar 
$T\bar{T}$ associated to $[4,0]$, we add the latter without enlarging 
the number of free parameters. The intersection of ${\sf d}_{1234}(\Delta_\varphi,\Delta_4)=0$ with   ${\sf d}_{1245}(\Delta_\varphi,\Delta_4)=0$ gives
$\Delta_\varphi\simeq-0.393$ and $\Delta_4\simeq3.666$. The exact value of the latter is 
$\Delta_4=\frac{18}5$.  Solving now the inhomogeneous system we find
${\sf p}_{\Delta_\varphi}\simeq-3.665$ to be compared with the exact result \cite{Cardy:1985yy}
\eq    
{\sf p}_{\Delta_\varphi}  =-
\frac{\Gamma(\frac65)^2\Gamma(\frac15)\Gamma(\frac25)}{\Gamma(\frac45)^3
\Gamma(\frac35)}\simeq-3.65312\,.
\en 
We can extract from ${\sf p}_{2,2}$ the estimate  $c\simeq-4.53$ of 
the central charge , while its exact value is $c=-\frac{22}5$.

\begin{figure}[tb]
\begin{center}
\includegraphics[width= 7 cm]{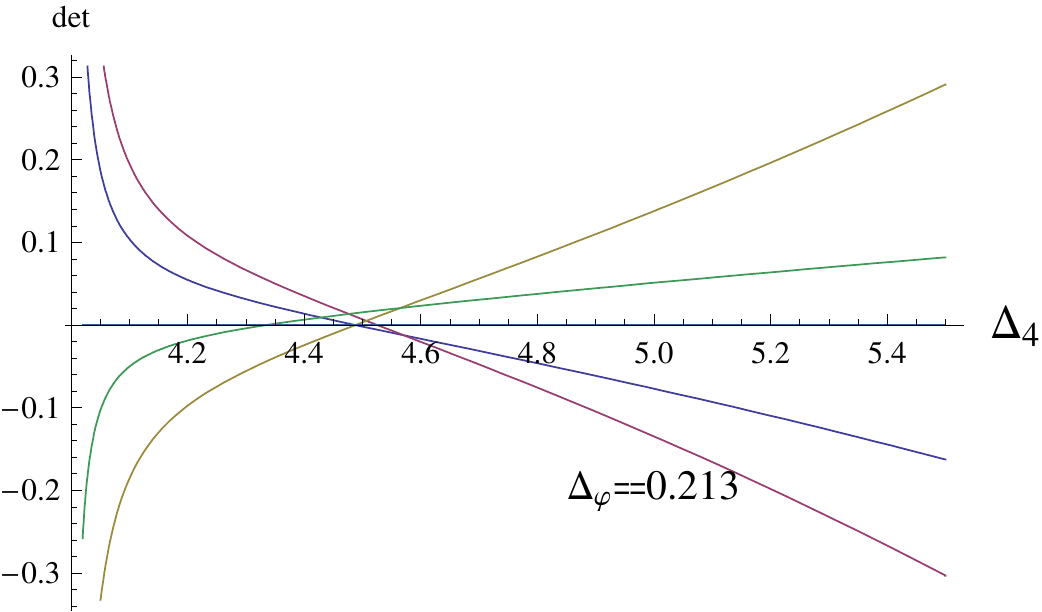}
\end{center}
\caption{Plot of some $3\times3$ minors around the solution (\ref{solution}) as functions of $\Delta_4$. Their convergence to zero near this solution supports 
our estimate of the critical exponent $\sigma$ for the three-dimensional Yang-Lee edge singularity.}
\label{Figure:2}  
\end{figure}

The last step is now to put $D=3$ in our formulae.
Entering in the three-dimensional world  we leave the golden eden of exactly solvable models and can resort solely to the 
internal consistency of the approach. 
Using the fusion rule truncated at the spin 4 operator we have only two free parameters. The intersection of the vanishing loci of the  $3\times3$ 
minors ${\sf d}_{123}(\Delta_\varphi,\Delta_4)$ and   
${\sf d}_{234}(\Delta_\varphi,\Delta_4)$ and the subsequent solution of the inhomogeneous system yield
\eq
\Delta_\varphi\simeq0.213,\Delta_4\simeq4.49,{\sf p}_{\Delta_\varphi}
\simeq-3.91,{\sf p}_{3,2}\simeq0.006
\label{solution}
\en 
where ${\sf p}_{\Delta_\varphi}  =\lambda_{\varphi\varphi\varphi}^2$ and 
 ${\sf p}_{3,2}  =\lambda_{\varphi\varphi T}^2$. 
Internal consistency requires that all the $3\times3$  minors made with the same equations (i.e. with the same indices $i=1,2,3,4$) should converge to zero 
or to small  values when approaching this solution. This is illustrated in Figure \ref{Figure:2}.  

\begin{figure}[tb]
\begin{center}
\includegraphics[width= 7 cm]{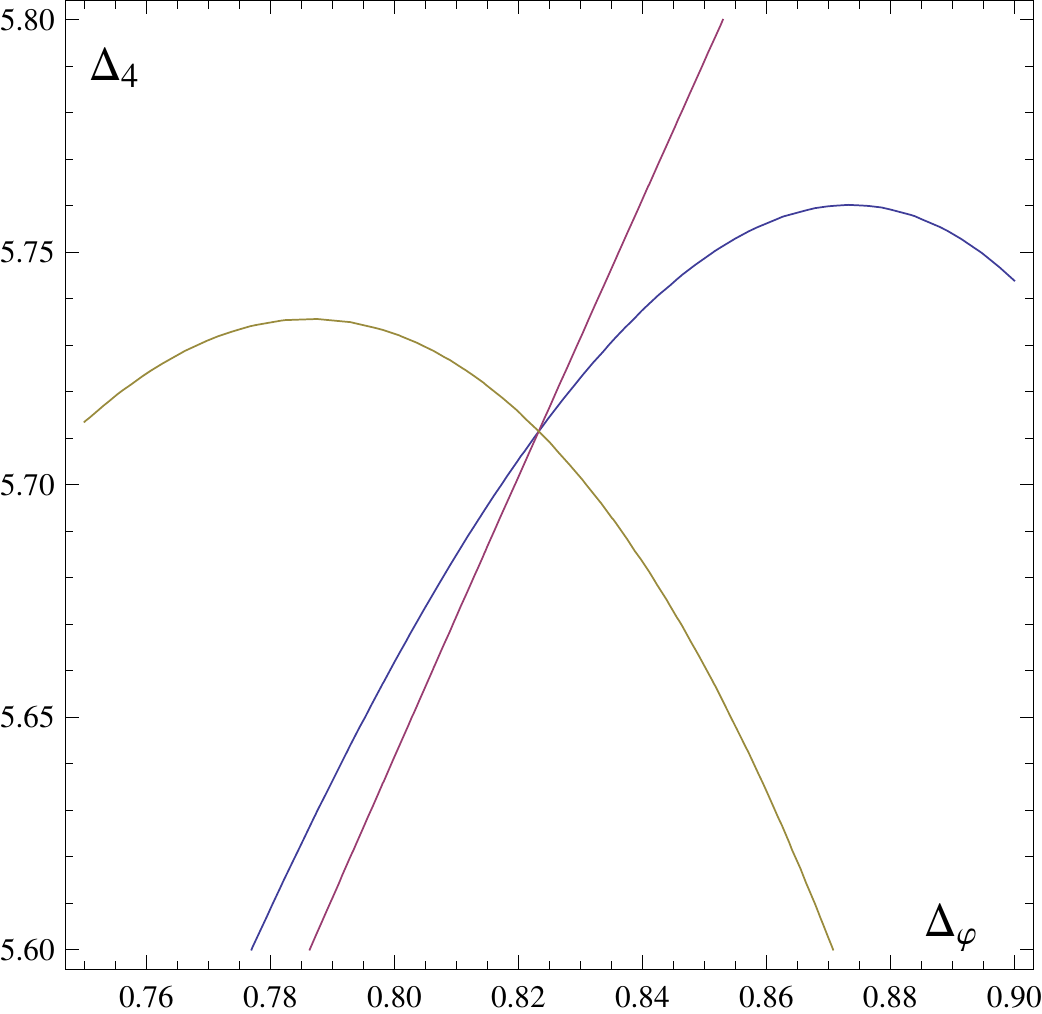}
\end{center}
\caption{Plot in the plane $(\Delta_\varphi,\Delta_4)$ of the zeros of some minors in the case of Yang-Lee edge singularity in four space dimensions.}
\label{Figure:3}  
\end{figure}

Inserting this value of $\Delta_{\varphi}$ in (\ref{sigma}) we obtain 
$\sigma\simeq0.076$, to be compared with the very accurate estimate 
\cite{Butera:2012tq} $\sigma=0.077(2)$, based on very long series expansions 
of dimer density in powers of the activity in a cubic lattice.
Similarly, with no more effort, we can apply the same analysis to the Yang Lee edge singularity in four space dimensions, obtaining
\eq
\Delta_\varphi\simeq0.823,\Delta_4\simeq5.71,{\sf p}_{\Delta_\varphi}
\simeq-2.86,{\sf p}_{4,2}\simeq0.40
\label{solution4}
\en 
and consequently $\sigma\simeq0.259$, to be compared with $\sigma=0.258(5)$ of
\cite{Butera:2012tq}.  Figure \ref{Figure:3} shows this solution as the common intersection of the zero loci of $3\times3$ minors.
 In principle the other physical parameters listed in (\ref{solution}) and 
(\ref{solution4}) could be checked with an $\epsilon=6-D$ expansion.

Clearly the present method may be successfully applied to other models. 
Note,  however, 
that in the fusion rules there are generally many more unknown 
scaling dimensions to be accounted for, which should require 
much larger determinants and, hence,  much larger orders of vanishing 
derivatives.

 For instance, in the 3d critical Ising model 
there are two more scalar primary operators of scaling dimension smaller 
than that of the spin 4 operator, one associated to $\varphi^4$ with 
$\Delta_{\varphi^4}=3.84(4)$, the other associated to   $\varphi^6$ with
$\Delta_{\varphi^6}=4.67(11)$ ($\Delta_4=5.0208(12)$ in this model).
A consistency check of the present method is to insert these values in the 
$5\times5$ minor associated  to the first 5 homogeneous equations as well as 
 to the primary operators appearing  in 
the fusion of $[\varphi]\times[\varphi]$, namely  
$[\varphi^2], [\varphi^4], [\varphi^6],[3,2],[\Delta_4,4]$.  Treating 
$\Delta_\varphi$ and $\Delta_{\varphi^2}$ as free parameters,
 the vanishing of this determinant
 yields the constraint $F(\Delta_\varphi,\Delta_{\varphi^2})=0$. This very constraint may be used 
to obtain a separate estimate 
of  $\Delta_\varphi$ and $\Delta_{\varphi^2}$, without resorting 
to higher derivatives. The key observation is that at this truncation level 
the fusion rule of 
$[\varphi]\times[\varphi]$ 
 coincides with that of $[\varphi^2]\times[\varphi^2]$, therefore in the latter case the constraint becomes $F(\Delta_{\varphi^2},\Delta_{\varphi^2})=0$, which has a 
discrete number of solutions.
One  is at $\Delta_{\varphi^2}\simeq1.447$. Inserting this value in the former constraint yields $\Delta_\varphi\simeq0.518$. The agreement with the most 
precise estimates \cite{{Campostrini:2002cf},{Hasenbusch:2011yya}}, 
$\Delta_\varphi=0.5182(2)$ and $\Delta_{\varphi^2}=1.4130(4)$ is even too good.
In order to have reliable results one should check their stability against the insertion of new operators. More information is needed on primary operators of higher scaling dimensions. Perhaps the recent progress on the knowledge of the  scaling properties of higher spin operators \cite{Komargodski:2012ek} 
could be very useful for this purpose.

In conclusion, we have reformulated the recently developed method of
implementing conformal bootstrap in diverse dimensions so that it
can be applied to a larger class of conformal field theories. Its
application to the Yang-Lee edge singularity  in three and four space 
dimensions as well as to the 3d critical  Ising model, gives rather
good results as compared to the best numerical methods.


\begin{thebibliography}{9}
\bibitem{rouche}
This is sometimes called the Rouch\'e -Capelli theorem.
\end{thebibliography}
\end{document}